\documentclass[epj-spec,final]{svjour}

\usepackage{graphicx}% Include figure files
\usepackage{dcolumn}% Align table columns on decimal point
\usepackage{bm}% bold math

\usepackage{amsmath,amssymb,graphicx,color}
\usepackage[numbers,sort&compress]{natbib}

\newcommand\diff{\mathrm{d}}

\newlength{\figwidth}
\begin{document}

% \preprint{APS/123-QED}

\title{The Localization Transition of the Two-Dimensional Lorentz Model}% Force line breaks with \\

\author{Teresa Bauer\inst{1} \and
Felix H\"of{}ling\inst{2} \fnmsep \inst{3} \and
Tobias Munk\inst{1} \and
Erwin Frey\inst{1} \and
Thomas Franosch\inst{4} \fnmsep \inst{1}}

\institute{%
Arnold Sommerfeld Center for Theoretical Physics (ASC) and Center for NanoScience (CeNS), Fakult{\"a}t f\"ur Physik,
Ludwig-Maximilians-Universit\"at M\"unchen, Theresienstra{\ss}e 37, 80333 M\"unchen, Germany
\and
Rudolf Peierls Centre for Theoretical Physics, 1~Keble Road, Oxford OX1 3NP, England, United Kingdom
\and
Max Planck Institute for Metals Research, Heisenbergstra{\ss}e 3, 70569 Stuttgart, Germany
\and
Institut f\"ur Theoretische Physik, Universit\"at Erlangen-N\"urnberg, Staudtstra{\ss}e 7, 91058 Erlangen, Germany
}%

\date{\today}% It is always \today, today,

\abstract{%
We investigate the dynamics of a single tracer particle performing Brownian motion
in a two-dimensional  course of randomly distributed hard obstacles.
At  a certain critical obstacle density,
the motion of the tracer becomes anomalous over many decades in time, which is rationalized in terms of an underlying percolation transition of the
void space. In the vicinity of this critical density the dynamics follows the anomalous one up to a crossover time scale where the motion
becomes either diffusive or localized.
We analyze the scaling behavior  of  the time-dependent diffusion coefficient $D(t)$ including corrections to scaling.
Away from the critical density, $D(t)$ exhibits universal hydrodynamic long-time tails both in the diffusive as well as in the localized phase.}

% \pacs{05.40.--a, 05.10.--a, 61.43--j, 64.60.Ht}
% {05.10.-a}{Computational methods in statistical physics and nonlinear dynamics}
% {05.40.--a}{Fluctuation phenomena, random processes, noise, and Brownian motion}
% {64.60.Ht}{Dynamic critical phenomena}
% {61.43.--j}{Disordered solids: amorphous solids, porous materials, fractals}

\maketitle
\section{\label{sec:intro}Introduction}
Understanding transport in  heterogeneous media is fundamental for a variety of applications ranging from
material sciences, porous catalysts, oil recovery, and even biological system. Often, the motion of particles inside
such  materials is strongly hindered due to the presence of slowly rearranging or immobilized obstacles of many different  length scales.
Many heterogeneous materials, e.g.
rocks, soils, cements, foams and ceramics,
consist of solid frames permeated by a  network of pores~\cite{Sahimi:Heterogeneous,Dagan:1984}, and
a mobile agent can meander  through  this static course of obstacles
and display long-range transport. Likewise,
transport in densely packed systems  is  strongly obstructed by the presence of surrounding particles
via their excluded volume effect.  In many cases a separation of time scales naturally occurs, for example
in strongly heterogeneous mixtures such as sodium ions in silicates~\cite{Meyer:2004,Kargl:2006,Voigtmann:2006} or size-disparate
soft or Yukawa spheres~\cite{Moreno:2006,Moreno:2006a,Kikuchi:2007},
leading to a  much  slower diffusion of one component.
Similarly, the dense packing of differently sized proteins, lipids and sugars in the cell cytoplasm leads to strongly suppressed transport known as
\emph{molecular crowding}~\cite{Ellis:2001,Ellis:2003,Tolic-Norrelykke:2004,Golding:2006}.
Again the motion of a smaller sized molecule is much faster than of  surrounding macromolecules and the small molecule
explores a quasi-static array of obstacles. Molecular crowding is also relevant in quasi two-dimensional systems such as protein diffusion in lipid bilayers
as studied by single molecule fluorescence microscopy~\cite{Deverall:2005} or fluorescence correlation spectroscopy~\cite{Kusumi:2005,Weiss:2003,Horton:2009}.

The motion of a tracer in these materials often displays anomalous transport as manifested in a subdiffusive increase of the mean-square displacement (MSD). 
This behavior
is displayed in  a finite window of time and a crossover to ordinary diffusion occurs at sufficiently long times. The exponent characterizing
 the subdiffusive behavior often appears to depend on the details of the system and even changes as the experimental parameters are varied.

The Lorentz model consisting of a single tracer exploring a course of randomly distributed frozen obstacles
constitutes a minimal model for transport in heterogeneous media~\cite{vanBeijeren:1982}. In its simplest variant the
obstacles are spheres or disks and the positions are Poisson distributed.
Hence, they may overlap and form clusters which restrict the motion of a pointlike tracer particle to the remaining void space.
Above a certain obstacle density, the void space no longer
permeates the entire system and a percolation transition occurs~\cite{Stauffer:Percolation}. It has been shown that
this transition is accompanied with a divergence of a characteristic length scale $\xi$, known as correlation length. The dynamics
close to the transition displays critical behavior and the mean-square displacement exhibits behavior similar to experimental
observations for heterogeneous media. It has been shown that the Lorentz model generically  leads to large crossover windows
explaining the apparent drift in characteristic
exponents~\cite{Lorentz_PRL:2006,Lorentz_JCP:2008,Hoefling:PhDthesis}. Recently, the two-dimensional Lorentz model has been introduced in the context
of lateral  diffusion of proteins in the plasma membrane~\cite{Sung:2006,Sung:2008a,Sung:2008}.

Systematic studies on the two-dimensional Lorentz model were mostly restricted to low densities focusing on the algebraic decay
 of the velocity autocorrelation functions (VACF)
and the non-analytic dependence of the diffusion coefficient on the obstacle
density~\cite{Bruin:1972,Bruin:1974,Alder:1978,Alder:1983,Alley:1979,Lowe:1993}. The recurrent collisions with
the obstacles lead to infinite memory resulting in a negative long-time tail $\sim -t^{-2}$ of the VACF~\cite{Weijland:1968,Ernst:1971a,vanBeijeren:1982};
yet close to the transition, the critical behavior shifts the onset of the hydrodynamic tail to longer and
longer times~\cite{Lorentz_LTT:2007}. The critical behavior of the Lorentz model in two-dimensions
is expected to be qualitatively different from the three-dimensional case since in the latter the conductances through
narrow channels determines the dynamic exponent from pure geometric reasons~\cite{Machta:1985,benAvraham:DiffusionInFractals}.
In the former the narrow gaps are expected to be less relevant and the universality of transport
on a percolating lattice should be recovered.  Beyond universality one would like to know
the range of validity of the universal behavior, the size of the crossover region, and the importance of corrections to scaling.

In this work we present simulation results for the two-dimensional Lorentz model for Brownian tracer particles, in particular for densities close to the
percolation transition. We have measured the mean-square displacement, the time-dependent diffusion coefficient, and the VACF, and analyze their respective
 critical behavior. Then
 we compare the  subdiffusive behavior as well as the diffusion coefficient with the predicted power-law behavior.
The non-algebraic decay of the VACF at long times  emerges  also for the case of a Brownian tracer corroborating the notion that
the frozen configuration space alone gives rise to persistent correlations in the dynamics.
A scaling theory that includes
the leading corrections to scaling is developed and tested against the simulation data by  suitable rectification plots.

\section{Lorentz Model}
The Lorentz model constitutes the minimal model for particle transport  through a disordered material. In its simplest
 variant, a single classical tracer particle traverses a $d$-dimensional array of frozen
hard-core obstacles of density $n$.
Each obstacle acts as a scattering center of radius $\sigma$ restricting the motion of the tracer to the void space.
For independently distributed scatterers the only control parameter characterizing the structure is then the dimensionless
number density $n^* = n \sigma^d$. Equivalently, one may employ the porosity $\varphi$, i.e., the volume fraction
accessible to the tracer due to the possibly overlapping obstacles. In the planar problem ($d=2$)
which we address in this work, one easily calculates
\begin{equation}
 \varphi = \exp(- \pi n \sigma^2)\, .
\end{equation}

Already at intermediate obstacle density, the void space decomposes into many pockets of different sizes, and long-range particle
transport occurs only through the void space that is percolating through the entire system.
At a certain obstacle density $n^*_c \approx 0.359$, the infinite component ceases to exist
and all particles are trapped in finite pockets. The problem of continuum percolation constitutes a critical phenomenon of purely
geometric origin~\cite{Stauffer:Percolation}, and a series of predictions has been made for the characteristic behavior in the close vicinity
 of the critical density $n_c^*$.  The linear dimension of the largest finite cluster (of the void space) defines
the correlation length $\xi$, which is expected to diverge as $\xi \sim | n^* -n_c^*|^{-\nu}$ as the critical density is approached. Below
the length scale $\xi$ the geometric structures appear, in a statistical sense, indistinguishable to the ones at $n_c^*$
forming the basis for the notion of self-similarity. Simultaneously, the root-mean-square size $\ell$ of all finite clusters
diverges, yet with a smaller exponent $\ell \sim |n_c-n_c^*|^{-\nu + \beta/2}$.
The same exponent $\beta$ governs the vanishing of the relative weight of the infinite
component as $n_c$ is increased towards the percolation threshold, $P_\infty \sim (n_c^*-n_c)^\beta$.
With respect to geometric properties, continuum percolation shares the same universality class as lattice percolation,
and in two dimensions the exact
values of the exponents are known  from a mapping to the Baxter line of the eight-vertex model $\nu = 4/3, \beta = 5/36 $~\cite{denNijs:1979,Nienhuis:1982,benAvraham:DiffusionInFractals}.

Transport of a single particle is expected to become anomalous and universal close to the percolation
thres\-hold independent of the details of the dynamics at the microscale. Here we consider a particle
undergoing Brownian motion confined to the void space. Then the short-time diffusion coefficient $D_0$ fixes the microscopic
time scale $t_0 := \sigma^2/D_0$, i.e., the typical time needed for the particle to diffuse the distance of one obstacle radius without obstruction.
The simplest quantity characterizing the motion of the tracer is the mean-square displacement
$\delta r^2(t) = \langle [\vec{R}(t)-\vec{R}(0)]^2\rangle$, where the brackets indicate averaging both over all initial positions
 of the particle as well as different realizations of the disorder. In particular, particles that are initially in a finite pocket
 will remain there forever and do not contribute to long-range transport.

Directly at the percolation threshold  ($n^*=n_c^*$) the void space is self-similar and the dynamics of a walker
becomes subdiffusive $\delta r^2(t) \sim t^{2/z}$ for long times $t\gg t_0$. The dynamic exponent $z$ is independent
of the geometric exponents of the percolation problem, but is determined from the universality class of random resistor networks.
For obstacle densities above  $n_c^*$,
all particles are trapped in finite clusters and correspondingly the mean-square displacement is expected to saturate at
the mean-square cluster size $\delta r^2(t\to \infty) = \ell^2$.
However, close to the transition the subdiffusive behavior should be visible in  a finite time window $t_0 \ll t \ll t_x$ where
$t_x$ denotes the crossover time to  localization. These arguments suggest that for small separation parameter $\epsilon := (n^*-n_c^*)/n_c^*$
 the mean-square displacement should obey the scaling law
\begin{equation}\label{eq:msd_scaling}
 \delta r^2(t;\epsilon) \simeq t^{2/z} \delta\hat{r}_\pm^2(\hat{t})\, , \qquad \hat{t}= t/t_x
\end{equation}
for  $\epsilon\!\downarrow\!0$ and $t\gg t_0$ and a scaling function $\delta\hat{r}^2_+(\cdot)$. We anticipate that scaling
is also obeyed on the diffusive side $(\epsilon\!\uparrow\!0)$ with a corresponding scaling function $\delta \hat{r}^2_-(\cdot)$ and the same
crossover time scale $t_x$. To describe the crossover from critical dynamics to localization/diffusion, the scaling functions
should exhibit the following asymptotics: $\delta \hat{r}^2_\pm(\hat{t} \to 0) = \textit{const.}$ and $\delta \hat{r}^2_+(\hat{t}\to \infty) \sim \hat{t}^{-2/z}$,
$\delta \hat{r}^2_-(\hat{t}\to \infty) \sim \hat{t}^{1-2/z}$, respectively. From the known long-time behavior on the localized side
one infers  for the crossover scaling time $t_x \sim \ell^z\sim |\epsilon|^{z (-\nu+\beta/2)}$. Interestingly, the relevant length scale that
determines the divergence of the inherited time is given by the mean cluster size $\ell$ rather than the correlation length $\xi$. Since $t_x$ also marks
the crossover to the diffusive regime for $\epsilon<0$, one immediately concludes that the long-time diffusion coefficient should vanish
 as $D \sim (-\epsilon)^\mu$ for $\epsilon \uparrow 0$ with
the conductivity exponent $\mu = (z-2) (\nu-\beta/2)$.  We use the value determined by Grassberger~\cite{Grassberger:1999} in
 high-precision computer simulations for the electrical conductivity of a percolating lattice, $\mu = 1.310\pm 0.001$, as reference value and
calculate the anomalous dimension to $z=3.036$. %\mu was determined in Grassberger, but called t
\section{Simulation Results}

\begin{figure}\sidecaption
\includegraphics[width=\figwidth]{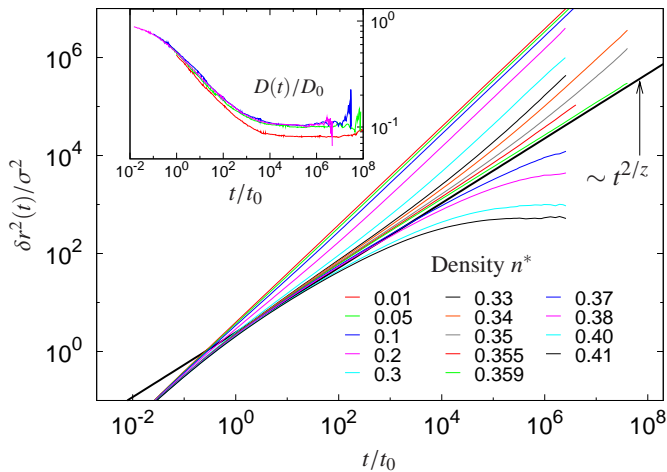}
\caption{Mean-square displacement $\delta r^2(t)$ of the Lorentz model for Brownian particles.
The obstacle density $n^*$ increases from top to bottom; the thick black line indicates the long-time
asymptote for anomalous transport at criticality $\delta r^2(t) \sim t^{2/z}$ with $z = 3.036$. The inset displays the
time-dependent diffusion coefficient $D(t)$ at obstacle density $n^* = 0.30$ for varying
algorithmic parameter $\tau_B/t_0 = 0.25 $, $0.015625 $, $ 0.0025 $, and $ 0.0005625$ .}
%\remark{change x-axis of inset: include shorter times (>1e-4?) and drop noise at long times (<1e-6)}
\label{fig:msd_brownian}
\end{figure}

We have performed Brownian dynamics simulations for a single particle moving in  a fixed array of hard disks of radius $\sigma$. The
obstacles are distributed independently with a fixed average density $n$. We have employed periodic boundary conditions for system
sizes of $L/\sigma = 10{,}000$ and generated Brownian trajectories for very long times following an algorithm
discussed in \cite{Scala:2007} which was also employed recently for the three-dimensional Lorentz model~\cite{Lorentz_JCP:2008}. In essence,
particles are propagated along a deterministic, straight trajectory with specular scattering every time
the tracer hits an obstacle, yet this dynamics is interrupted at regular time intervals $\tau_B$, where new velocities are
drawn from a Maxwell distribution of variance~$v^2$. Then on time scales longer compared to  $\tau_B$ and length scales larger
than $v \tau_B$, a free particle undergoes Brownian motion with diffusion coefficient $D_0 = v^2 \tau_B/4$. In the presence of the
obstacles  the particle can still be considered as Brownian walker at the microscale with short-time diffusion coefficient $D_0$, provided
 $\tau_B$ is small relative to the inverse collision rate $\tau_c=1/2\pi n \sigma v$.
We shall use the characteristic time $t_0 = \sigma^2/D_0$ as our basic unit of time, i.e., the time a free particle needs to
traverse an obstacles radius. The algorithmic condition
to mimic Brownian dynamics at the microscale is thus given by $\tau_B /t_0 \ll 1/ (2n^*)^2$.
We have verified that $\tau_B/t_0 =0.0025$ is sufficiently small in order that the long-time behavior is independent
of the microparameters $v^2$ and $ \tau_B$ as exemplified in the inset of Fig.~\ref{fig:msd_brownian} for a moderate
obstacle density $n^* = 0.3$. There the time-dependent diffusion coefficient is displayed for different $\tau_B$ and the
 curves neatly superimpose  for $\tau_B < 0.015625 t_0$. For the production runs we have fixed $\tau_B = 0.0025 t_0$ and have
calculated mean-square displacements as running-time averages over several trajectories for at least 100 different realizations of the disorder resulting 
in more than 775 trajectories in total for each density $n^*$  to generate high-accuracy data. The trajectories extend over huge time
 windows of typically $2.5 \times 10^6 t_0$, yet close to the percolation threshold  and for densities $n^* \leq 0.15$ 15 times longer trajectories  
 have used.
 With current computing resources, a single trajectory at the critical density and for the longest simulation times runs
approximately 40 hours on one core of a Quad Core Intel(R) Xeon(R) CPU X5365 (3.00\,GHz).

Results for the mean-square displacement $\delta r^2(t)$
for all obstacle densities are exhibited in Fig.~\ref{fig:msd_brownian} on double logarithmic scales.
First one should notice, that the data display almost no noise even for the longest times. For short times, all data start from
 the short-time diffusive motion, $\delta r^2(t) = 4 D_0 t$, and only at times  $t \simeq t_0$ the presence of the obstacles suppresses the
 motion and the curves  fan out.
For low obstacle densities, the long-time behavior is again diffusive yet with a suppressed diffusion coeffficient~$D(n^*)$.
 On the other hand, the mean-square displacements for high $n^*$ saturate at long times,
 reflecting the fact that all particles are localized.

The localized and diffusive curves are nicely discriminated by a critical density $n_c^* = 0.359$ where the MSD behaves subdiffusively
 over at least six decades in time, i.e., it extends to  our longest  observation times. This critical density coincides with the
 numerical estimate $n_c^*=0.359072(4)$ for continuum percolation~\cite{Quintanilla:2000,Quintanilla:2007}. 
The value $z = 3.036$,
as inferred from the exponent $\mu$ determined by finite-size scaling of the conductivity at the critical point~\cite{Grassberger:1999} provides an excellent
description of the long-time behavior of the critical MSD $\delta r^2(t) \sim t^{2/z}$.
Our simulations provide the first quantitative  test  that the 2d Lorentz model shares the universality class of two-dimensional random resistor networks.

For densities close to $n_c^*$, the data follow the critical one up to some  finite  time where they cross over to either
 diffusive or localized behavior. By naked eye one infers already that
this crossover time increases as the critical density is gradually approached. Let us mention that in the three-dimensional
 Lorentz model, the curves off the critical point deviate much more
from the critical one, than in the planar Lorentz model. Nevertheless  they still display subdiffusion in a finite time window, yet with apparent
 density-dependent exponents~\cite{Lorentz_PRL:2006,Lorentz_JCP:2008}.

\subsection{Time-dependent diffusion coefficients}

A quantity more sensitive  to the anomalous transport behavior is given in terms of the time-dependent diffusion coefficient
\begin{equation}
 D(t) := \frac{1}{2d} \frac{\diff }{\diff t} \delta r^2(t) \, ,
\end{equation}
where the dimension is $d=2$ for the planar problem. We have taken numerical derivatives of the MSDs taking advantage
 of the fact that $\delta r^2(t)$ varies significantly only on logarithmic scales. Since the MSDs are calculated using
 our standard blocking scheme~\cite{Glassy_GPU:2010}, the numerical derivatives essentially do not introduce new noise to the data.
\begin{figure}\sidecaption
\includegraphics[width=\figwidth]{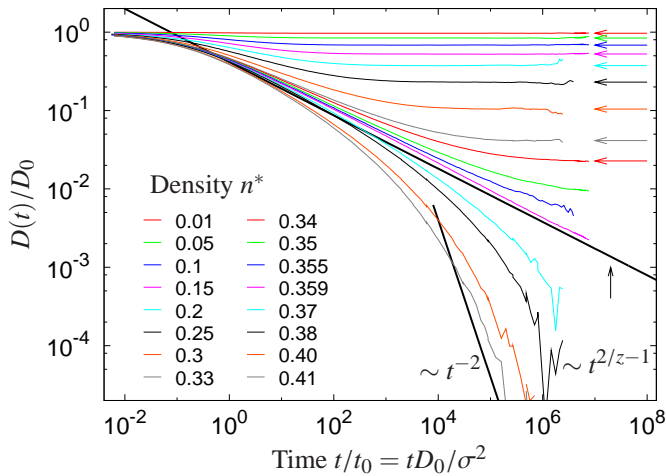}
\caption{The time-dependent diffusion coefficient $D(t) := (1/4) \diff \delta r^2(t)/\diff t$. The obstacle density $n^*$ increases
 from top to bottom; the arrows indicate the long-time diffusion coefficient $D$. For the critical density, the time-dependent
 diffusion coefficient vanishes as a power law,
$D(t) \sim t^{2/z-1}$. The thick line indicates the power law $t^{-2}$ expected as long-time asymptote in the localized regime due to the cul-de-sacs.  }
\label{fig:time_dependent_diffusion}
\end{figure}
The time-dependent diffusion coefficient $D(t)$ is displayed in Fig.~\ref{fig:time_dependent_diffusion} for the same densities
 considered above. First, one notices  that all curves start
from the short-time diffusion constant $D_0$, corroborating that our numerical algorithm reproduces Brownian motion at
small time and length scales.  For increasing time
 $D(t)$  is gradually suppressed reflecting that obstacles can only slow down the overdamped dynamics. For densities below $n_c^*$     the time-dependent diffusion coefficient
approaches a nonzero limit $D$ for long-times. The values of the long-time diffusion constant $D$ decrease rapidly as the critical density is approached from below.
Directly at the critical point, $D(t)$ reaches a power-law long-time asymptote $D(t) \sim t^{2/z-1}$
 corresponding to a subdiffusive mean-square displacement.
For densities above the critical one, the time-dependent diffusion coefficient vanishes even more rapidly.
%long-time decay $\ell^2/2 td $ much slower than observed for the $D(t)$ considered here.
Following the argument of persistent correlations due to power-law distributed
exit rates of the cul-de-sacs, one should expect a universal long-time tail $D(t) \sim t^{-2}$ in the entire localized
phase~\cite{Machta:1985}. Such a behavior has indeed been observed recently in molecular dynamics
 simulations for the two-dimensional Lorentz model for ballistic particles~\cite{Lorentz_LTT:2007}, though they considered
 the velocity autocorrelation function rather than $D(t)$. Our simulations  exhibit clear evidence that this tail
 remains present for Brownian particles too as we shall argue below.

\begin{figure}[t]\sidecaption
\includegraphics[width=\figwidth]{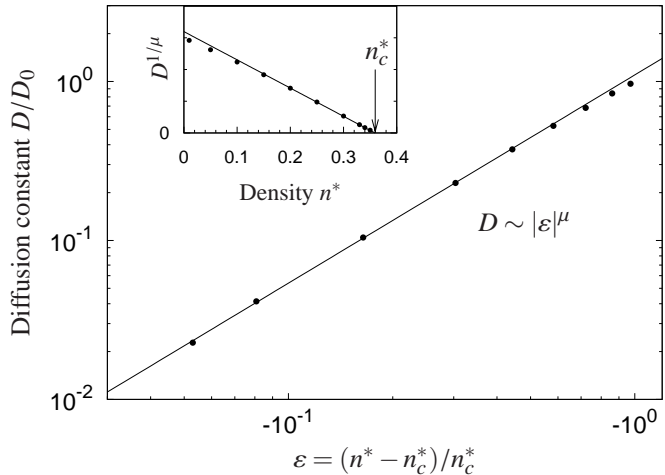}
\caption{Scaling behavior of the long-time diffusion coefficient $D$ with separation parameter $\epsilon = (n^*-n_c^*)/n_c^*$.  The straight line in the rectification plot (inset)
confirms the value of the conductivity exponent $\mu = 1.309$.
}
\label{fig:diff_coeff}
\end{figure}
The diffusion constants $D$ extracted as  long-time limits of $D(t)$ are displayed in Fig.~\ref{fig:diff_coeff} for varying
obstacle density. Over the investigated range of densities,
the diffusion constant is suppressed by a factor of $100$.
It vanishes as the critical density is approached and follows amazingly well the scaling prediction $D\sim (-\epsilon)^\mu$.
Even for the lowest density considered, where the motion is practically unobstructed by the obstacles, $D(n^* = 0.01) = 0.97 D_0$ deviates
by only $8.4 \%$ from the scaling asymptote.
It appears  as a coincidence that the critical regime connects down to the low-density asymptote without an intermediate region of moderate obstructed motion.
 For the corresponding three-dimensional system the convergence towards the scaling behavior is
approached slowlier, however, since the corresponding conductivity exponent is much higher $\mu_{3\text{d}} = 2.88$, the diffusion vanishes much more rapidly
 and a suppression by five orders of magnitude
can be observed~\cite{Lorentz_JCP:2008}. The rectification plot in the inset of Fig.~\ref{fig:diff_coeff} corroborates that  $\mu= 1.310$ obtained
by measuring the conductivity on a lattice close to percolation is indeed the correct value. Our simulations for Brownian particles provide an independent test
  that two-dimensional random resistor networks and the planar Lorentz model indeed share the same universality class. The critical density has
been determined by extrapolating
the straight line in the rectification plot to zero diffusivity, yielding $n_c^* =0.359\pm 0.001$. The
thus determined value of $n_c^*$  was used throughout this work to simulate the critical dynamics.

\subsection{Velocity autocorrelation functions}

Let us also discuss the velocity autocorrelation function (VACF),  $Z(t) = \langle \vec{v}(t) \cdot \vec{v}(0) \rangle/d$,
for the Brownian particle. Although
the notion of velocity for Brownian particles is conceptually questionable,  their correlation function is well defined for
times $t>0$. Here we rely again on numerical derivatives, i.e.,
we employ
\begin{equation}
 Z(t) := \frac{1}{2d} \frac{\diff^2}{\diff t^2} \delta r^2(t)
\end{equation}
as definition. Then the relation to the time-dependent diffusion coefficient is provided by
\begin{equation}\label{eq:Green_Kubo}
 D(t) = D_0 +  \int_{0^+}^t Z(t')\, \diff t' \, ,
\end{equation}
where the integral is evaluated excluding the time $t=0$. This form constitutes the analog of the Green-Kubo
 relation, alternatively one can include a $\delta$-distribution in the VACF
 to account for the Brownian motion at the microscale.

\begin{figure}\sidecaption
\includegraphics[width=\figwidth]{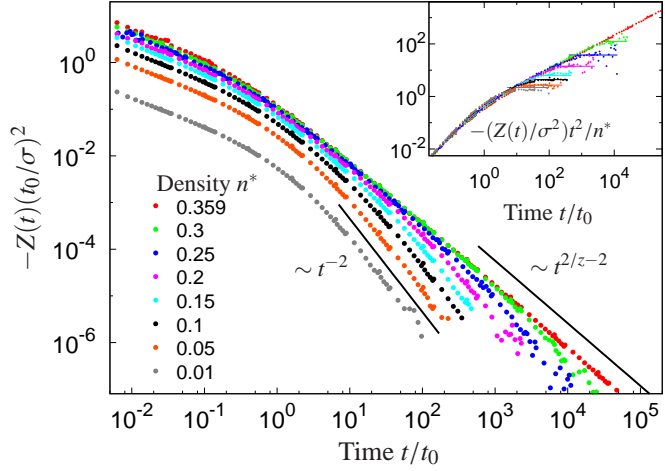}
\caption{Velocity autocorrelation function $Z(t) =  (1/4) \diff^2 \delta r^2(t)/\diff t^2$ for Brownian particles
in the planar Lorentz model. The thick lines indicate the hydrodynamic tail $t^{-2}$ and the critical behavior $t^{2/z-2}$, respectively.
 The inset displays a rectification plot $-t^2 Z(t)/\sigma^2/n^*$ as a function of time. }
\label{fig:long_time_tail}
\end{figure}
Figure~\ref{fig:long_time_tail} displays the VACF, and one first observes that it is negative for all times, except on time scales
associated with our algorithmic microparameter $\tau_B$.
This fact is consistent with the notion that obstacles can only slow down the diffusion, Eq.~(\ref{eq:Green_Kubo}), and in the case of Brownian motion
one can show that the VACF is a completely monotone function, see Appendix \ref{sec:Bernstein}.
 The
long-time behavior for the diffusive regime ($n<n_c)$
is characterized by persistent correlations that slowly decay as a power-law. The low densities display a
 tail $Z(t) \sim - t^{-2}$ consistent with the theoretical prediction for kinetic theory for ballistic particles~\cite{Ernst:1971a}.
It  has been anticipated earlier~\cite{vanBeijeren:1982} that also Brownian particles exhibit the same behavior,
since the long-time correlations originate from repeated encounters of the same frozen heterogeneities. Indeed the Lorentz model for Brownian tracers can be solved
analytically to lowest order in the scattering density $n^*$ and the time-dependence  of the VACF including its long-time tail can be worked out
exactly~\cite{Lorentz_VACF:2010}.
Nevertheless, to the best of our knowledge,
our simulation results provide the first direct evidence for this universality at all densities. As the density is gradually increased, the overall
signal in the VACF becomes larger and the exponent of the power-law tail appears to drift.
A rectification plot shows that the $t^{-2}$ behavior is assumed for all densities as the late-time relaxation. The critical
asymptote appears in an  intermediate  window which extends to longer and longer  times as the critical density is approached.
Our data also show that the amplitude of the hydrodynamic tail diverges close to $n_c^*$ which can be
rationalized using the same arguments as in the ballistic case~\cite{Lorentz_LTT:2007}.

\section{Dynamic scaling analysis}

\begin{figure}\sidecaption
\includegraphics[width=\figwidth]{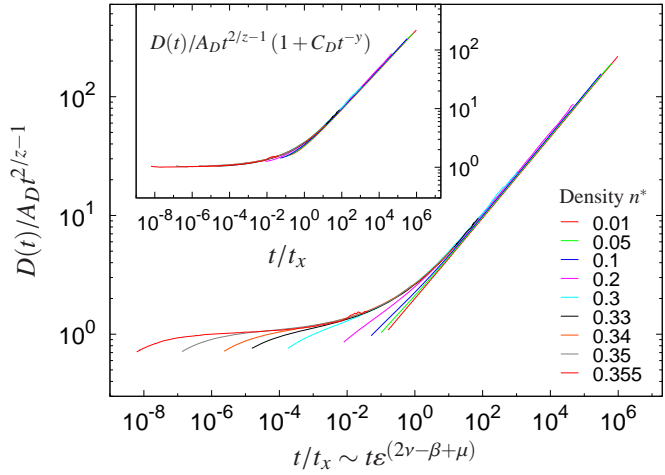}
\caption{Scaling behavior of the time-dependent diffusion coefficient for the densities below $n_c^*$. The scaling variable reads $t_x = t_0 |\epsilon|^{\beta-2\nu-\mu}$
The inset includes the leading correction to scaling with a correction amplitude $C_D = -0.14 t_0^y$ consistent with the critical density.
}
\label{fig:D_scaling}
\end{figure}
The power-law behavior in both the mean-square displacement or  the time-dependent diffusion coefficient at criticality and
the vanishing of the diffusion constant as a power law upon approaching the critical density is merely one aspect of the
 critical behavior. Yet, the universality hypothesis suggests a much more
sensitive test in terms of scaling. For example, the mean-square displacement $\delta r^2(t;\epsilon)$ for small
separation parameters $|\epsilon| \ll 1$ and long times $t\gg t_0$ is expected to fulfill  Eq.~(\ref{eq:msd_scaling}). Here we show that
 the time-dependent diffusion coefficient
$D(t;\epsilon)$   can be used equivalently to test the scaling prediction. Taking derivatives, we suggest
\begin{equation}
 D(t;\epsilon) \simeq t^{2/z-1} \hat{D}_\pm(\hat{t}) \, ,\qquad \hat{t} = t/t_x
\end{equation}
with the scaling time $t_x := t_0 |\epsilon|^{-(2\nu -\beta + \mu)}$. The connection with the scaling function for the mean-square displacement is easily established,
\begin{equation}\label{eq:D_scaling}
 \hat{D}_\pm(\hat{t}) = \frac{1}{2d} \left[ \frac{2}{z} \delta \hat{r}^2_\pm(\hat{t}) + \hat{t} \frac{\diff}{\diff \hat{t}} \delta \hat{r}^2(\hat{t}) \right]\, .
\end{equation}
For short rescaled times, $\hat{D}_\pm(\hat{t}\to 0) = \textit{const.} =: A_D$ and the critical behavior is recovered. For long times $\hat{D}_-(\hat{t}) \sim \hat{t}^{1-2/z}$ such that
diffusion is reached for $\epsilon<0$. On the localized side,
the mean-square displacements saturate and the leading
behavior $\delta r^2_+(\hat{t}) \sim \hat{t}^{-2/z}$ cancels exactly in Eq.~\ref{eq:D_scaling}, thus $\hat{D}_+(\hat{t}) = o(\hat{t}^{-2/z})$.

\begin{figure}\sidecaption
\includegraphics[width=\figwidth]{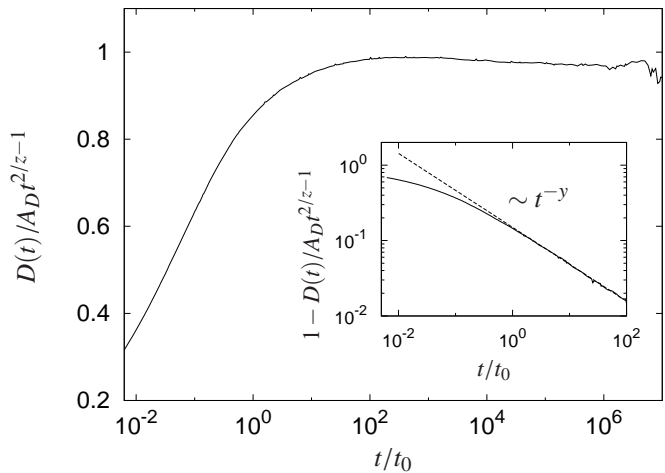}
\caption{Rectification of the time-dependent diffusion coefficient $ D(t)$ at the critical density $n_c^*$. The critical amplitude
is determined to $A_D=  0.508 D_0 t_0^{1-2/z}$. The inset displays the approach of the MSD towards the critical law. The thick
 line indicates a power laws with the universal correction exponent $y= 0.49$.}
\label{fig:diffusion_critical}
\end{figure}

Figure~\ref{fig:D_scaling} displays a rectification plot for $D(t)$ for obstacle densities below the critical ones. For large rescaled
times the curves nicely superimpose, though this reflects merely the fact that the diffusion regime is reached for all cases and that the
long-time diffusion coefficient obeys the scaling prediction $D \sim (-\epsilon)^\mu$. For short rescaled times the curves converge to a
constant which is given by the long-time behavior of the time-dependent diffusion coefficient at the critical point. The fanning out of the
curves arises due to corrections to scaling and eventually because of the crossover to the microscopic regime.

To gain further insight into the scaling behavior we extend our scaling hypothesis by a generic irrelevant scaling variable. Then it has
been shown recently within a cluster-resolved scaling theory~\cite{Percolation_EPL:2008} that the mean-square displacement should obey
\begin{equation}\label{eq:msd_correct}
 \delta r^2(t;\epsilon) = t^{2/z} \delta \hat{r}_\pm^2(\hat{t}) \left[ 1 + t^{-y} \Delta_\pm(\hat{t}) \right] \, ,
\end{equation}
where $y$ is another universal exponent characterizing the approach of the critical dynamical behavior. It is connected to a correction-to-scaling
exponent $\Omega$ for the cluster-size distribution via the scaling relation $y = \Omega (\nu d - \beta) / [z (\nu-\beta/2)]$. For the two-dimensional
case the value $y =  0.49(3)$~\cite{Percolation_EPL:2008} was determined for
random walks on a lattice, which we shall use in the following.
Taking derivatives with respect to time, the corresponding prediction for the time-dependent diffusion coefficient is
\begin{equation}\label{eq:diff_correct}
 D(t;\epsilon) \simeq t^{2/z-1} \hat{D}_\pm(\hat{t}) \left[ 1 + t^{-y} \Delta_\pm^D(\hat{t}) \right] \, ,
\end{equation}
 where the new scaling function $\Delta_\pm^D(\hat{t})$ is connected to the one of the mean-square displacement via
\begin{align}
 2d \hat{D}_\pm(\hat{t}) \Delta_\pm^D(\hat{t}) =&  \left(\frac{2}{z}-y \right) \delta \hat{r}^2_\pm(\hat{t}) \Delta_\pm(\hat{t}) \nonumber \\
&+ \hat{t} \frac{\diff}{\diff \hat{t}} \left[
\delta \hat{r}^2_\pm(\hat{t}) \Delta_\pm(\hat{t}) \right]  \, .
\end{align}
For small rescaled times the correction-to-scaling function reduces to a constant $\Delta_\pm^D(\hat{t}\to 0) = \textit{const.} =: C_D$.
and one easily infers the relation to the correction-to-scaling constant $C := \Delta_\pm(\hat{t}\to 0)$ for the mean-square
displacement: $C_D = C (2- yz)/2$. In particular at criticality, the time-dependent diffusion coefficient displays a
power-law correction for long times
\begin{equation}\label{eq:diffusion_correction}
 D(t;\epsilon =0) \simeq A_D t^{2/z-1} \left[ 1 + C_D t^{-y} \right] \, .
\end{equation}
For long rescaled times $\hat{t}\to \infty$,  the correction to scaling function behaves as a power-law
again $\Delta_-^D(\hat{t}\to \infty) \simeq \tilde{C}_\pm^D \hat{t}^y$, yielding corrections for the asymptotic behavior
of the long-time diffusion constant
\begin{equation}\label{eq:diff_constant_correction}
 D(\epsilon) \sim (-\epsilon)^\mu \left[ 1 + \tilde{C}_-^D t_0^{-y} (-\epsilon)^{y \mu z /(z-2)} \right] \, .
\end{equation}

The time-dependent diffusion coefficient $D(t)$ at the critical density is displayed in Fig.~\ref{fig:diffusion_critical} in a rectification plot.
Within the statistical errors of our simulation one observes a saturation at long times provided the established value for $z$ is used, indicating that the
asymptotic behavior is reached. At very long times the curve starts to deviate again
due to statistical fluctuations, possibly finite-size effects, and the uncertainty of the value for the critical obstacle density.
The amplitude of the critical relaxation $A_D = \lim_{t\to \infty} t^{1-2/z} D(t;\epsilon =0)$ has been determined to 
$A_D = 0.508 D_0 t_0^{1-2/z}$ from our numerical data to optimize data collapse for the scaling in the diffusive and localized regime, see below. 
The approach towards this power-law behavior is consistent
with a power law according to Eq.~(\ref{eq:diffusion_correction})
with a correction to scaling amplitude $C_D = -0.14 t_0^y$, see inset of Fig.~\ref{fig:diffusion_critical}.

It appears that for a deeper analysis of the correction to scaling behavior as suggested by Eqs.~(\ref{eq:msd_correct}) and
(\ref{eq:diff_correct}), the precise form of the scaling
function $\Delta_\pm^D(\hat{t})$ has to be known.
From general arguments, we know that the correction function for the MSD, $\Delta_\pm(\hat{t})$, smoothly interpolates between
 a constant $\Delta_\pm(\hat{t}\to 0) =: C$ and power laws $\Delta_\pm(\hat{t}\to \infty) \simeq \tilde{C}_\pm \hat{t}^y$. Then the
correction term for the time-dependent diffusion coefficient behaves for large rescaled times $\hat{t}\to \infty$ as
 $\hat{D}(\hat{t}) \Delta_-^D(\hat{t}) \sim \hat{t}^{y-2/z}$ on the diffusive side and
  $\hat{D}(\hat{t}) \Delta_-^D(\hat{t}) =  o(\hat{t}^{y-2/z})$ in the localized regime.
Upon closer inspection of the correction behavior, Eq.~(\ref{eq:diff_constant_correction}), and  the simulation
results, Fig.~\ref{fig:diff_coeff}, the corrections
are very small for large rescaled times. Then one may approximate $\Delta_\pm^D(\hat{t})$ by its short-time asymptote
$\Delta_\pm^D(\hat{t}) = C_D$ for all rescaled times, i.e., ignore the corrections on the diffusive/localized time regime.
The result of this procedure is displayed in the
 inset of Fig.~\ref{fig:D_scaling} for the diffusive side with a quite impressive improvement of the
scaling behavior. 
Let us emphasize that the only new parameter $C_D$ is in principle fixed by the critical
 behavior, such that no free parameters enter the scaling plot. 
 \begin{figure}\sidecaption
\includegraphics[width=\figwidth]{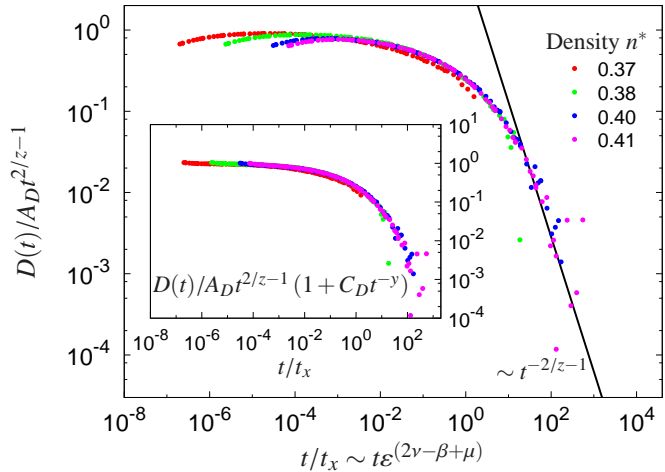}
\caption{Scaling behavior of the time-dependent diffusion coefficient for the densities above $n_c^*$ with  
the crossover time   $t_x = t_0 |\epsilon|^{\beta-2\nu-\mu}$. 
The thick line indicates the power law  $t^{-2/z-1}$ which is expected to hold in the entire localized phase.
The inset includes the leading correction to scaling with the same  correction amplitude $C_D = -0.14 t_0^y$ as in the diffusive regime. 
}
\label{fig:D_scaling_loc}
\end{figure}

The scaling behavior for the time-dependent diffusion coefficient $D(t)$ is tested for the localized regime, $n^* > n_c^*$, in Fig.~\ref{fig:D_scaling_loc}. Again the curves approach a constant at short rescaled times as the critical density is approached. Accounting for the correction by the same procedure as for the diffusive side yields an almost perfect data collapse without any new parameters.
The scaling behavior at large times reflects that also the approach of the saturation in the MSDs is universal. From
our general discussion, the scaling function is known to vanish rapidly $\hat{D}_+(\hat{t}) = o(\hat{t}^{-2/z})$. Following Machta and Moore~\cite{Machta:1985} there should universal
power-law correlations in the entire localized phase, $D(t)\sim t^{-2}$ for $t \to \infty$, due to the meandering of the particle in the self-similar cul-de-sacs. Assuming that the crossover from the critical law to these universal hydrodynamic tails
is again given by $t_x$ as we have argued earlier for ballistic particles~\cite{Lorentz_LTT:2007}, leads to a scaling prediction of $\hat{D}_+(\hat{t}) \sim \hat{t}^{-2/z-1}$.
In particular, one expects a divergence of the prefactor of the tail in the localized phase according to $t_x^{2/z+1}$.
 This prediction is indicated in Fig.~\ref{fig:D_scaling_loc} and provides a nice description of the data.

\section{Conclusion}

The dynamics of a tracer particle in a densely packed planar course of obstacles  has been investigated by Brownian dynamics simulation. The slowing down of the dynamics
close to the percolation threshold is accompanied by critical behavior observed over more that 6 decades in time. We corroborate that the planar Lorentz model shares a universality class with the random resistor network  where the critical exponents are known from earlier simulations.  We have shown that the time-dependent diffusion coefficient constitutes a suitable quantity to analyze the scaling behavior close to the transition. The corresponding scaling relations have been derived and extended by the leading correction.
We find that scaling behavior is in general well obeyed and the corrections to scaling appear much less important than for the three-dimensional case~\cite{Lorentz_JCP:2008}.

The Lorentz model exhibits power-law long-time anomalies away from the critical density due to repeated encounters with the same scatterer. These tails have been derived originally for ballistic particles, yet they turn out to be universal irrespective of the dynamics at microscopic scales. Then the velocity autocorrelation, defined via a second derivative of the mean-square displacement exhibits the tails even for Brownian tracers. On the localized side we also find long-time tails due to the self-similar distribution of exit times of the cul-de-sacs~\cite{Machta:1985}, again irrespective of the microscopic dynamics. Interestingly, these long-time tails are part of the scaling function for the time-dependent diffusion coefficient.

The assumption that the obstacles are distributed independently is certainly an oversimplification in real systems.
Then one would like to extend the Lorentz model
where the matrix consists of some frozen-in configuration of a strongly interacting liquid or a snapshot of a slowly rearranging matrix of obstacles. Second,
experiments are usually for a finite concentration of particles meandering in the array of obstacles and one may ask at what time and length scales these
interaction of the tracers modifies the dynamics in the labyrinth.
In three dimensions
an intriguing interplay of the physics of the glassy dynamics and the localization transition has been discovered recently~\cite{Meyer:2004,Kargl:2006,Voigtmann:2006,Moreno:2006,Moreno:2006a,Kikuchi:2007}, and since the glass transition in two-dimensions is qualitatively similar~\cite{Santen:2000,Bayer:2007} one may hope that the physics of the planar Lorentz model is applicable in size-disparate two-dimensional mixtures.

\begin{acknowledgement}
Financial support  from the Deutsche Forschungsgemeinschaft via contract No. FR 850/6-1 and from the Konrad-Adenauer-Stiftung (T.B.) is gratefully acknowledged.
This project is supported by the  German Excellence Initiative via the program ``Nanosystems Initiative Munich (NIM).''
\end{acknowledgement}

\appendix

\section{Appendix: Completely monotone functions}\label{sec:Bernstein}
In this Appendix we develop a spectral representation for the mean-square displacement and the velocity-autocorrelation function for arbitrary dimension $d$.

For a Brownian particle in an external potential $U(\vec{r})$ the time-evolution of the conditional probability distribution $\Psi(\vec{r},t)$ to find
 the particle at $\vec{r}$ at time $t$ provided it has been at $\vec{r}'$ and some earlier time $t'$ is governed by the Smoluchowski equation
\begin{equation}\label{eq:Smoluchowski}
\partial_t \Psi(\vec{r},t| \vec{r}' t') =  \frac{\partial}{\partial \vec{r}} \left[ \frac{D_0}{k_B T} \frac{\partial U}{\partial \vec{r}} \Psi \right] + D_0\frac{\partial^2 \Psi}{\partial \vec{r}^2} \equiv \hat{\Omega}(\vec{r}) \Psi\, ,
\end{equation}
where  $\hat{\Omega}(\vec{r})$ denotes the Smoluchowski operator acting on the position $\vec{r}$. At the very end, we are interested
in hard potentials with infinite barriers, however we anticipate that this case is assumed as limit of smooth potentials becoming increasingly steep.

One can also consider the evolution of $\Psi$ with respect to the conditional time and
one can show that
\begin{equation}
-\partial_{t'} \Psi(\vec{r},t| \vec{r}' t') = - \frac{D_0}{k_B T} \frac{\partial U}{\partial \vec{r}'}
 \frac{\partial \Psi }{\partial \vec{r}'} + D_0\frac{\partial^2 \Psi}{\partial \vec{r}'{}^2} \equiv \hat{\Omega}^+(\vec{r}') \Psi\, ,
\end{equation}
where the adjoint $\hat{\Omega}^+(\vec{r}')$ is with respect to the standard scalar product.  Furthermore, $\hat{\Omega}^+(\vec{r}')$ is
 identified with the  backward Smoluchowski operator and  now acts on $\vec{r}'$.

The mean-square displacement is obtained as an average
\begin{align}
 \delta r^2(t-t') &\equiv \int\!\! \diff \vec{r} \diff \vec{r}'\,  (\vec{r}-\vec{r}')^2 \Psi(\vec{r}t |  \vec{r}' t') \Psi_{\text{eq}}(\vec{r}') \, ,
\end{align}
where $\Psi_{\text{eq}}(\vec{r}) = Z^{-1} \exp(-U(\vec{r})/k_B T)$ denotes the equilibrium distribution.
In principle one may also introduce a disorder average for different realizations of the potential $U(\vec{r})$, yet we
anticipate that for large enough systems the quantities of interest
are self-averaging.  Since in equilibrium the MSD is stationary, it depends only on the time difference and one derives
\begin{align}
& \frac{\diff^2}{\diff t^2} \delta r^2(t-t') = - \frac{\diff}{\diff t}\frac{\diff}{\diff t'} \delta r^2(t-t') \nonumber \\
&= \int\!\! \diff \vec{r} \diff \vec{r}' \,(\vec{r}-\vec{r}')^2 \left[ \hat{\Omega}(\vec{r}) \hat{\Omega}^+(\vec{r}')
 \Psi(\vec{r}t |  \vec{r}' t')\right] \Psi_{\text{eq}}(\vec{r}')
\nonumber \\
&=  \int\!\! \diff \vec{r} \diff \vec{r}'    \Psi(\vec{r}t |  \vec{r}' t')
\left[ \hat{\Omega}(\vec{r}')\hat{\Omega}^+(\vec{r})  (\vec{r}-\vec{r}')^2 \Psi_{\text{eq}}(\vec{r}') \right] \nonumber \\
&=  \int\!\!\diff \vec{r} \diff \vec{r}'\,    \Psi(\vec{r}t |  \vec{r}' t')
\left[ \hat{\Omega}^+(\vec{r}')\hat{\Omega}^+(\vec{r})  (\vec{r}-\vec{r}')^2  \right] \Psi_{\text{eq}}(\vec{r}') \nonumber \\
&= -2  \int\!\! \diff \vec{r} \diff \vec{r}'  \, \left[ \hat{\Omega}^+(\vec{r}) \vec{r} \right]\cdot
 \left[ \hat{\Omega}^+(\vec{r}') \vec{r}' \right] \Psi(\vec{r}t |  \vec{r}' t')
 \Psi_{\text{eq}}(\vec{r}') \, ,
\end{align}
where in the second to last line the property of the Smoluchowski operator $ \hat{\Omega}(\vec{r}) \left[ A(\vec{r}) \Psi_{\text{eq}}(\vec{r}) \right] =
\left[ \hat{\Omega}^+(\vec{r})  A(\vec{r})  \right] \Psi_{\text{eq}}(\vec{r})$
valid for any well-behaved function $A(\vec{r})$ has been employed. The preceding result shows that the second derivative of the MSD
can be interpreted essentially as the negative of the autocorrelation function of $\hat{\Omega}^+(\vec{r}) \vec{r}$. To make connection
 with the ballistic case it is helpful to
introduce $\vec{v} = \text{i} \hat{\Omega}^+(\vec{r}) \vec{r}$ as a formal velocity, and one recovers the usual relation to the velocity autocorrelation function
\begin{equation}
\langle \vec{v}(t) \cdot \vec{v}(t') \rangle = \frac{1}{2} \frac{\diff^2}{\diff t^2} \delta r^2(t-t') \, .
\end{equation}

Next we recall that autocorrelation functions $C(t) = \langle A(t)^* A(0) \rangle$   for overdamped dynamics
are completely monotone, i.e., their derivatives exhibit fixed sign
\begin{equation}
(-1)^n \frac{\diff^n}{\diff t^n} C(t) \geq 0\quad \text{for all } n\in \mathbb{N}_0\, ,\quad  t\geq 0 \, .
\end{equation}
A sketch of  a non-rigorous proof is as follows. First, consider the complex scalar product
\begin{equation}
 \langle A | B \rangle = \int \diff \vec{r} A(\vec{r})^* B(\vec{r}) \Psi_{\text{eq}}(\vec{r}) \, .
\end{equation}
Then one easily verifies that $\hat{\Omega}^+$ is hermitian with respect to this scalar product $\langle A | \Omega^+ B \rangle = \langle \Omega^+ A | B \rangle$.
Then with the formal solution of Eq.~(\ref{eq:Smoluchowski}),
$\Psi(\vec{r}t| \vec{r}'t') = \exp[ (t-t') \hat{\Omega}(\vec{r}) ] \delta(\vec{r}-\vec{r}')$, one finds a representation of the autocorrelation function as
\begin{equation}
 C(t) = \langle \text{e}^{\Omega^+ t }A | A \rangle \, .
\end{equation}
Yet since $\hat{\Omega}^+$ is hermitian all eigenvalues are real, and by a `variational principle',
\begin{equation}
 \langle \hat{\Omega}^+ A | A \rangle = - D_0 \int \diff \vec{r} \left| \frac{\partial A(\vec{r})}{\partial \vec{r}}  \right|^2 \Psi_{\text{eq}}(\vec{r}) \leq 0 \, ,
\end{equation}
 also negative semi-definite. Zero constitutes the non-degenerate eigenvalue with constant eigenfunction $|0\rangle$.
A spectral decomposition of the backwards Smoluchowski operator $\hat{\Omega}^+ =
- \sum_\lambda \lambda |\lambda \rangle  \langle \lambda | $ in terms of eigenfunctions
$\hat{\Omega}^+ |\lambda \rangle = \lambda | \lambda \rangle$ shows that an autocorrelation
function can be  represented as
\begin{equation}
 C(t) = \sum_\lambda \left| \langle A| \lambda \rangle \right|^2 \exp(-\lambda t) \, , \qquad \text{for } t>0 \, .
\end{equation}
From this representation one immediately infers that $C(t)$ is completely monotone. By the famous
Bernstein theorem~\cite{Feller:ProbabilityTheory} the converse is also true, i.e.,
any completely monotone function allows for a representation as a superposition of relaxing exponentials with positive weights.

For the VACF one concludes
\begin{equation}
 \langle \vec{v}(t) \cdot \vec{v}(t') \rangle = -  \sum_\lambda \left| \langle \lambda | \hat{\Omega}^+ \vec{r} \rangle \right|^2  \exp(-\lambda t)\, .
\end{equation}
Since the equilibrium state $| 0\rangle$ is annihilated by the backward Smoluchowski operator $\hat{\Omega}^+ | 0 \rangle = 0$, the sum extends in
fact only over positive eigenvalues $\lambda> 0$.

Integration yields the time dependent diffusion coefficient
\begin{equation}
 D(t) = D_0  - \sum_{\lambda>0} \left| \langle \lambda | \hat{\Omega}^+ \vec{r} \rangle \right|^2 \frac{1-\exp(-\lambda t)}{ \lambda d } \, ,
\end{equation}
and one immediately infers that $D(t)$ is monotonically decreasing to the long-time diffusion coefficient
\begin{equation}
 D = D_0  - \sum_{\lambda>0} \left| \langle \lambda | \hat{\Omega}^+ \vec{r} \rangle \right|^2 \frac{1}{ \lambda d } \, .
\end{equation}
For the mean-square displacement one obtains the representation
\begin{equation}
 \delta r^2(t) = 2 d D_0 t - 2\sum_{\lambda>0} \left| \langle \lambda | \hat{\Omega}^+ \vec{r} \rangle \right|^2 \frac{\lambda t - 1 + \exp(-\lambda t)}{ \lambda^2} \, ,
\end{equation}
valid for $t\geq 0$.
\bibliographystyle{apsrev}
\bibliography{lorentzgas}

\end{document}